\documentclass[useAMS,usenatbib]{emulateapj}
\newcommand\msun {M$_{\odot}$}
\def\approxgt{\ifmmode \rlap{$>$}{}_{{}_{{}_{\textstyle\sim}}} \else%
$\rlap{$>$}{}_{{}_{{}_{\textstyle\sim}}}$\fi} 
\def\approxlt{\ifmmode \rlap{$<$}{}_{{}_{{}_{\textstyle\sim}}} \else%
$\rlap{$<$}{}_{{}_{{}_{\textstyle\sim}}}$\fi}

\def\arcmin{\hbox{$^\prime$}}
\def\arcsec{\hbox{$^{\prime\prime}$}}

\def\flx{erg cm$^{-2}$ s$^{-1}$}
\def\lum{erg s$^{-1}$}
\def\xmm{{\it XMM--Newton}}
\def\chan{{\it Chandra}}
\def\src{NGC~3921}

\shorttitle{The bright ULX in NGC~3921}
\shortauthors{Jonker et al.}

\begin{document}

\title{The nature of the bright ULX X--2 in NGC~3921: a \chan\, position and HST candidate counterpart}

\author{P.G.~Jonker\altaffilmark{1,2,3}} 

\email{p.jonker@sron.nl}
\author{M.~Heida\altaffilmark{1,2}}
\author{M.A.P.~Torres\altaffilmark{1,3}}
\author{J.C.A.~Miller--Jones\altaffilmark{4}}
\author{A.C.~Fabian\altaffilmark{5}}
\author{E.M.~Ratti\altaffilmark{1}}
\author{G.~Miniutti\altaffilmark{6}}
\author{D.J.~Walton\altaffilmark{5}}
\author{T.P.~Roberts\altaffilmark{7}}

\altaffiltext{1}{SRON, Netherlands Institute for Space Research, Sorbonnelaan 2,
  3584~CA, Utrecht, The Netherlands}
\altaffiltext{2}{Department of Astrophysics/IMAPP, Radboud University Nijmegen,
P.O.~Box 9010, 6500 GL, Nijmegen, The Netherlands}
\altaffiltext{3}{Harvard--Smithsonian  Center for Astrophysics, 60 Garden Street, Cambridge, MA~02138, U.S.A.}
\altaffiltext{4}{International Centre for Radio Astronomy Research, Curtin University, GPO Box U1987, Perth, WA 6845, Australia}
\altaffiltext{5}{Institute of Astronomy, Madingley Road, Cambridge CB3 0HA, United Kingdom}
\altaffiltext{6}{Centro de Astrobiolog\'{i}a (CSIC--INTA), Dep.~de Astrof\'{i}sica;
ESA, P.O.~Box 78, E-28691 Villanueva de la Ca\~nada, Madrid, Spain}
\altaffiltext{7}{Department of Physics, Durham University, South Road, Durham DH1 3LE, United Kingdom}

\begin{abstract} \noindent We report on \chan\, observations of the bright ultra--luminous X--ray (ULX)
  source in NGC~3921. Previous \xmm\, observations reported in the literature
  show the presence of a bright ULX at a 0.5--10 keV luminosity of 2$\times
  10^{40}$\lum. Our \chan\, observation finds the source at a lower luminosity
  of $\approx 8\times10^{39}$\lum, furthermore, we provide a \chan\, position of
  the ULX accurate to 0.7\arcsec\, at 90\% confidence. The X--ray variability
  makes it unlikely that the high luminosity is caused by several separate 
  X--ray sources. In 3 epochs of archival Hubble Space Telescope (HST)
  observations we find  a candidate counterpart to the ULX. There is direct
  evidence for  variability between the two epochs of WFPC2 F814W observations
  with  the observation obtained in 2000 showing a brighter source. 
  Furthermore, converting the 1994 F336W and 2000 F300W WFPC2 and the 2010 F336W
  WFC3 observations to the Johnson $U$--band filter assuming a spectral type of
  O7~I we find evidence for a  brightening of the $U$--band light in 2000. Using
  the higher  resolution WFC3 observations the candidate counterpart is
  resolved  into two sources of similar color. We discuss the  nature of the ULX
  and the probable association with the optical  counterpart(s). Finally, we
  investigate a potential new explanation  for some (bright) ULXs as the
  decaying stages of flares caused by  the tidal disruption of a star by a
  recoiled supermassive black  hole. However, we find that there should be at
  most only 1 of such systems within z=0.08.

\end{abstract}


\section{Introduction} 

Ultraluminous X-ray sources (ULXs) are off-nuclear X-ray point sources in
nearby galaxies with X-ray luminosities, L${_X}\approxgt 1\times 10^{39}$ --
$10^{42}$ erg s$^{-1}$ (\citealt{1999ApJ...519...89C};
\citealt{2009Natur.460...73F}). Their X-ray luminosities are suggestive of
intermediate-mass (10$^2$--10$^5$ \msun) black holes (IMBHs) if they radiate
isotropically at sub-Eddington levels as observed in stellar black holes and
Active Galactic Nucleii (AGNs). Hence, ULXs could be a new kind of black
hole with masses in--between the stellar--mass black holes found in X--ray
binaries and the supermassive black holes (SMBHs; $\approxgt 1\times 10^6$
\msun) found in the centers of galaxies. These IMBH could be the building
blocks of SMBHs (e.g.~\citealt{2010Natur.466.1049V}). For a comprehensive
overview of ULXs see \citet{2011NewAR..55..166F}.

Recent studies of the luminosity functions of ULXs and fainter
extra-galactic X--ray binaries showed evidence for a break at a
luminosity $\sim2\times 10^{40}$\lum\,(\citealt{2011ApJ...741...49S};
\citealt{2012MNRAS.419.2095M}). This, together with the identification
of an ultra--luminous spectral state under super-Eddington accretion
rarely observed in Galactic black hole X-ray binaries (e.g.,
\citealt{2009MNRAS.397.1836G}), led to the idea that the majority of
ULXs are stellar--mass black holes. These stellar--mass black holes
have particular radiation mechanisms, including beaming effects
(\citealt{2001ApJ...552L.109K}) and/or truly super--Eddington emission
(\citealt{2002ApJ...568L..97B}). Nevertheless, a number of the
brightest ULXs with an X--ray luminosity $\approxgt 2\times
10^{40}$\lum\, remains hard to explain as X--ray binaries. The higher
the luminosity, the less likely it is that they can be explained as
the high--luminosity end of the X--ray binary luminosity function.

A seemingly convincing case for an IMBH is provided by the variable, very bright ULX
ESO~243--49 X--1 (\citealt{2009Natur.460...73F}). Its X--ray luminosity is too high for a
stellar--mass black hole even in the presence of some beaming. Given the evidence for the
detection of a redshifted H$\alpha$ emission line in the optical counterpart to ESO~243--49
X--1 (\citealt{2010ApJ...721L.102W}), the uncertainty on its distance is reduced with respect
to other bright ULXs. Another case for an IMBH is M82~X41.4+60, whereas it does not reach peak
luminosities as high as ESO~243--49 X--1, the maximum luminosity is still uncomfortably high
for a stellar--mass black hole (\citealt{2003ApJ...586L..61S}).  \citet{2009ApJ...696.1712F}
showed that M82~X41.4+60 displays state changes similar to those of stellar--mass black holes
when at sub--Eddington accretion rates. However, differences in interpretation exist. The
temperature of the thermal emission is $> 1$ keV, which is high for an IMBH.
\citet{2006ApJ...652L.105O} describe the spectrum as that from a slim disk obtaining a black
hole mass of 20--30\,\msun.  Similarly, \citet{2009PASJ...61S.263M} interprete the thermal
emission as coming from a thick corona and they find a black hole mass of $<$200 \msun.

A further interesting alternative explanation for some (bright) ULXs
is that the ULX is caused by a recoiling SMBH in the few
million years after the recoil event (\citealt{2010MNRAS.407..645J}
and references therein). This scenario is not feasible for galaxies
where there is evidence for the presence of a central supermassive
black hole from e.g.~an AGN. The reason is that the recoiled SMBH
needs to be replaced by a new SMBH in a subsequent merger, and the
time for that is longer than the few million years lifetime of the
recoiled SMBH as a ULX.

The potential explanation of the ULXs with luminosities above the
break in the luminosity function as IMBHs or recoiled SMBHs warrant
further investigation of these sources. So far, less than a dozen
sources with ${\rm L_X \gtrsim 10^{41}}$\lum\, have been found
(i.e.~in the Cartwheel galaxy, \citealt{2006MNRAS.373.1627W};
M~82~X--1, \citealt{2009ApJ...696.1712F} and
\citealt{2003ApJ...586L..61S}; ESO~243-49,
\citealt{2009Natur.460...73F}; in NGC~5775,
\citealt{2008MNRAS.390...59L}; and in CXO~J122518.6+144545,
\citealt{2010MNRAS.407..645J}). Eight more bright off--nuclear X--ray
sources, of which 2 have ${\rm L_X \gtrsim 10^{41}}$\lum, were
recently reported by \citet{2012MNRAS.423.1154S}, who suggest that their
properties are consistent with accreting IMBHs in the hard state
(cf.~\citealt{2011MNRAS.416.1844W}).

Here, we present \chan, archival Very Large Array (VLA), and archival
HST/WFPC2 and HST/WFC3 observations of the bright ULX in \src\, first
found by \citet{2004MNRAS.353..221N} called NGC~3921~X--2. \src\, is
classified as a protoelliptical that was formed during a major merger
taking place approximately 0.7 Gyr ago, where one of the merging
galaxies was gas rich and the other gas poor
(\citealt{1996AJ....111..109S}; \citealt{1996AJ....112.1839S}). Given
that \src\, has an AGN the recoiling black hole scenario desribed in
\citet{2010MNRAS.407..645J} is not applicable for the ULX under study
in this galaxy. We adopt $\Omega_m= 0.3$, $\Omega_\Lambda=0.7$, and
$H_0 = 70$ km s$^{-1}$ Mpc$^{-1}$ to convert the redshift of \src\,to
a distance measurement.

\section{Observations, analysis and results}

\subsection{{\it Chandra} X-ray observation} 

We have obtained a 6.0~ks--long observation with the \chan\, X--ray
observatory (\citealt{2002PASP..114....1W}) covering the \xmm\, error
circle of the bright ULX in \src~(\citealt{2004MNRAS.353..221N}) on
August 14, 2011 (MJD 55787.89362).  The \chan\, observation has been
performed using the S3 CCD of the Advanced CCD Imaging Spectrometer
(ACIS) detector (\citealt{1997AAS...190.3404G}; ACIS--S). The
observation identification (ID) number for the data presented here is
13296.  We reprocessed and analyzed the data using the {\sc CIAO 4.3}
software developed by the \chan\, X--ray Center and employing {\sc
  CALDB} version 4.3. The data telemetry mode was set to {\it very
  faint}, which allows for a thorough rejection of events caused by
cosmic rays.

Since, by design, the \xmm\, source position falls near the optical
axis of the telescope, the size of the point spread function is smaller
than the ACIS pixel size. Therefore, we follow the method of
\citet{2004ApJ...610.1204L} implemented in the {\sc CIAO 4.3} tool
{\sc acis\_process\_events} to improve the image quality of the ACIS
data.

Using {\sc wavdetect} we have identified the source related to the AGN
in \src~(cf.~\citealt{2004MNRAS.353..221N}). Using our accurate radio
position of the AGN (see below) we apply a boresight correction to the
\chan\, observation of size 0.155\arcsec\, to the Right Ascencion (RA)
and -0.079\arcsec\, to the Declination (Dec). For this we used the
{\sc wcsupdate} tool inside {\sc CIAO}. We checked the position of the
AGN after the boresight correction and it matches the radio position
to within 0.02\arcsec. However, as the radio position is accurate to
$\sim 0.1$\arcsec\, we take the latter as the residual 1$-\sigma$
boresight uncertainty. As the boresight correction involves the
position of one source only we can not account for effects of
rotation.

In the boresight corrected X--ray image we detect an 8--count X--ray
source in the \xmm\, error circle at a position of RA = 11:51:07.723
and Dec=+55:04:14.98. In decimal degrees the position, stating
in brackets the formal 1 sigma {\sc wavdetect} uncertainty in
localising the source on the ACIS S3 CCD, is (RA,Dec)=177.78218(4),
+55.07083(3). The 90 per cent uncertainty in the source position is
(RA,Dec)=(0.3\arcsec,0.24\arcsec), with a total uncertainty of
0.4\arcsec.  This is slightly lower than the uncertainty of
0.5\arcsec\, (90 per cent confidence) for low count sources found by
\citet{2005ApJ...635..907H}, which could potentially be due to the use
of the sub-pixel event repositioning that can currently be
applied. For the best position we, however, conservatively use an
uncertainty of 0.5\arcsec\, for the uncertainty in localizing the
source on the CCD.  Given the possibility that the residual boresight
correction is due to a systematic effect we assume, conservatively,
that the overall error on the \chan\, position of the ULX in \src\, is
determined by adding the localization uncertainty and the boresight
uncertainty (0.2\arcsec\, at 90 per cent confidence). Overall, the
\chan\, 90 per cent uncertainty on the source position is
0.7\arcsec. The source is detected 0.3\arcmin\, off-axis on the ACIS
S3 CCD.

In order to verify that the source we detect in the \xmm\, error
circle is not due to residuals from a cosmic ray event, we investigated
the arrival times of the X-ray photons from the source. Indeed, the
arrival times of the X-ray photons do not cluster in time as one would
expect for residual emission due to a cosmic ray event.

We also detect 4 counts from the source that
\citet{2004MNRAS.353..221N} list as \src~X--3. The best \chan\,
position is (RA,Dec)=(11:51:03.53,+55:04:50.2) which corresponds to
177.76472(5), 55.08061(6) in decimal degrees, where the digit in
between brackets denotes the 1~$\sigma$ {\sc wavdetect} uncertainty in
localising the source on the CCD. We do not detect \src~X--4 that was
also considered a ULX by \citet{2004MNRAS.353..221N}.  Possibly, the
source is variable as well and it was below our detection limit.
Alternatively, the detection of \src~X--4 was spurious resulting from
the incomplete modeling of the \xmm\, point spread function
(cf.~\citealt{2011A&A...534A..34R}). It was excluded from the
catalogue from \citet{2011MNRAS.416.1844W} for that reason.

We have selected a circular region of 6 pixel ($\approx$3\arcsec) radius
centered on the source position of \src~~X--2 to extract the source
counts in the energy range of 0.3--8 keV. We limited the radius to
exclude as much as possible the central parts of the galaxy, where
extended emission either due to hot gas or due to unresolved faint point
sources may be present.  Similarly, we have used a circular region with
a radius of 80 pixels away from any source but on the same S3 ACIS CCD
to extract background counts. The net number of background subtracted
source counts is 7.7. From the number of background events and the
background area we find that the predicted number of background source
photons is less than 0.3. For a model of a power law with index 1.7 and
negligible interstellar absorption (\citealt{1990ARA&A..28..215D} give
$N_H=1.1\times 10^{20}$ cm$^{-2}$ in the direction of \src) this yields
a 0.3--8 keV flux of 1$\times 10^{-14}$\flx. However, given the low
number of detected counts, the uncertainty on this value is about a
factor of 2.

Using standard cosmological parameters the redshift converts to a
distance of 80 Mpc for NGC~3921, which makes the X-ray luminosity in the
range 0.3--8 keV ${\mathrm L_X\approx 8\times 10^{39}}$\lum. For
comparison the 0.5--10 keV luminosity is the same when assuming a
powerlaw spectral model with an index of 1.7 and the Galactic
interstellar extinction given above. 

\subsection{{\it HST} WFPC2 observations }

We have analysed archival Hubble Space Telescope (HST) data obtained
with the Wide Field Planetary Camera 2 (WFPC2) in the F336W, the F555W
and the F814W filters corresponding roughly to the Johnson $U$-, $V$-
and Cousins $I$-bands, respectively. We converted the WFPC2 {\sc
  vegamag} to the Johnson $U$--, $V$-- and Cousins $I$--band
magnitudes using the {\sc synphot} package running under {\sc stsdas}
in {\sc iraf} assuming the spectral energy distribution of the O7~I
star BZ~64. We use this procedure and spectral type for all filter
transformations after this. The HST data were obtained on June 28,
1994 (MJD 49623) and the proposal identification number is 5416. Using
these HST observations, \citet{1996AJ....112.1839S} provide the $V$--
and $I$--band magnitude (see their table 3 source number 129) of a
source consistent with our \chan\, X--ray position of
NGC~3921~X--2. As we will see below, the $V$ and $I$--band magnitudes
of $V$=24.03 and $I$=24.02 determined by \citet{1996AJ....112.1839S}
are fully consistent with the values we derive below using the same
HST data. We summed the reduced drizzled images provided by the Space
Telescope Science Institute giving total exposure times of
2$\times$600~s, 2$\times$1200~s plus 30~s, and 2$\times$900~s plus
30~s for the $U$-, $V$- and $I$-bands, respectively.

\begin{figure*} 
\includegraphics[angle=270,width=16cm,clip]{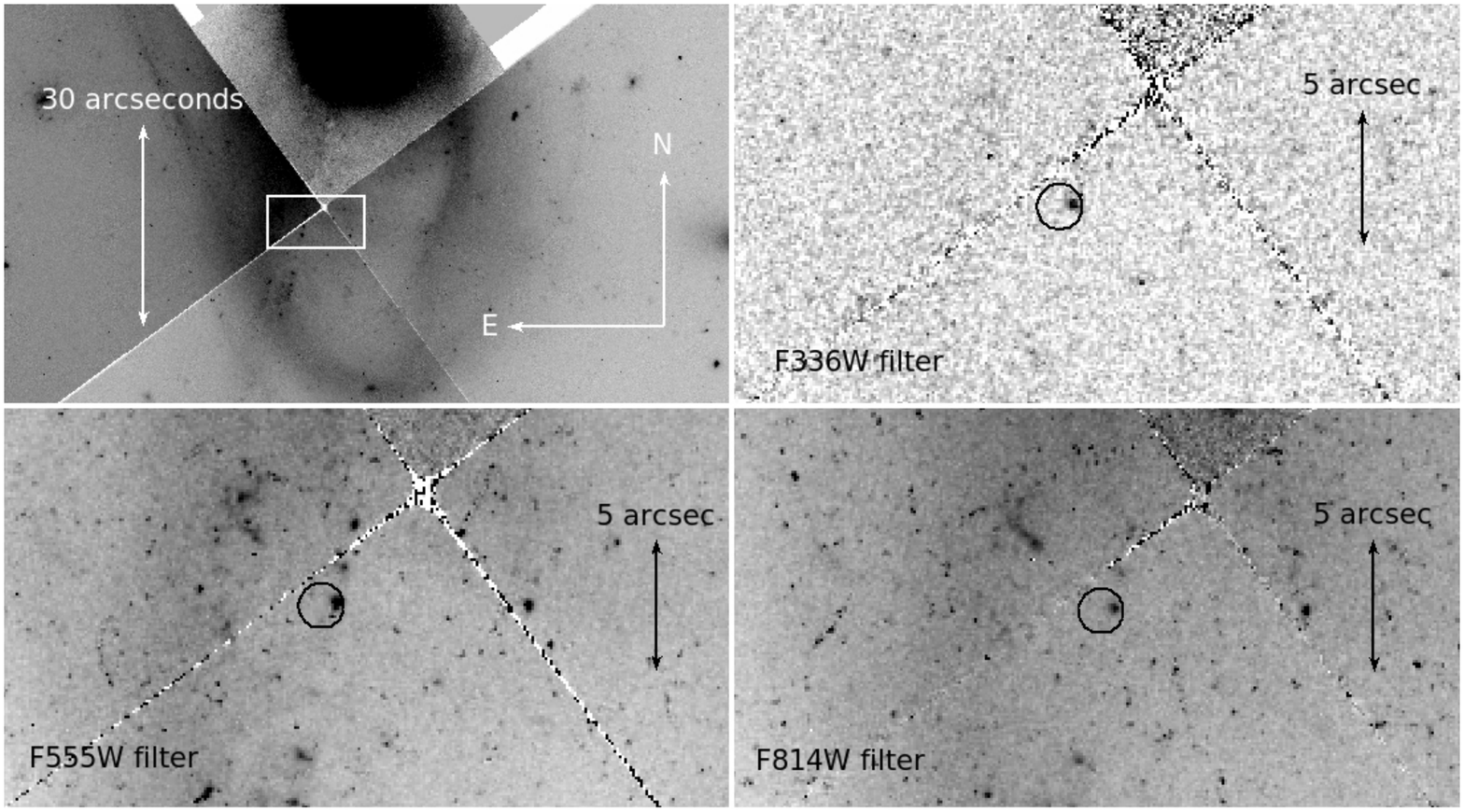}
\caption{All the images are made using the 1994 WFPC2
  observations. The small circle with 0.85\arcsec\, radius present in
  three sub--plots represents the 90\% confidence astrometric error
  circle on the position of the ULX. North is up and East is to the left in all panels. {\it Top left panel:} Overview of
  the field of \src\,in the F555W filter. The box approximates the
  size of the small zoomed images. {\it Top right panel:} The
  combined, 1200~s--long, HST WFPC2 F336W image ($U$--band), showing
  the blue source near the edge of the small circle.  {\it Bottom left
    panel:} Zoom-in of the combined 2430~s WFPC2 F555W ($V$-band)
  image around the position of the bright ULX in \src. {\it Bottom
    right panel:} Zoom-in of the combined 1830~s WFPC2 F814W
  ($I$-band) image around the position of the bright ULX in \src. }
\label{ima} \end{figure*}

We performed astrometry within ${\sc midas}$, comparing the positions
of stars in the field of view of the WFPC2 against entries from the
seventh release of the Sloan Digital Sky Survey (SDSS DR7) $r^\prime$
catalogue (\citealt{2009yCat.2294....0A}). An astrometric solution was
computed by fitting for the reference point position, the scale and
the position angle, considering all the sources that are not saturated
and appear stellar and unblended. We obtain a solution with
0.12\arcsec\, root-mean-square (rms) residuals from 15 stars well
distributed on the image. The astrometric accuracy of SDSS DR7 is
better than 75 milli-arcseconds rms per coordinate, plus 20-30
milli-arcseconds due to systematic errors
(\citealt{2003AJ....125.1559P}). For the accuracy on our stellar
positions we sum in quadrature the residuals of the astrometry with
the rms accuracy of the catalogue and add linearly the systematic
uncertainty on the SDSS positions. The resulting positional accuracy
with respect to the International Celestial Reference System (ICRS) is
0.17\arcsec\, on both RA and Dec.~at 1~$\sigma$.  The combined \chan\,
and HST WFPC2 positional accuracy at 90\% confidence is therefore
0.85\arcsec (0.7\arcsec\, comes from tying the \chan\, astrometric
frame to the ICRS, see above, and an additional uncertainty of
0.48\arcsec\, from tying the HST frame to the ICRS, adding both in
quadrature).

There is a bright point source present inside the 90\% confidence
astrometric error circle in all three bands at a position
RA=11:51:07.65 and Dec=+55:04:15.2 ([RA,Dec]=177.78187(5),
+55.07089(5) in decimal degrees, where the value in the between
brackets denotes the 1--$\sigma$ uncertainty; see Fig.~\ref{ima}).  We
used sky values determined local to the source of interest as the sky
background varies substantially due to a variable amount of unresolved
stars of the galaxy. We performed point-spread function fitting for
the photometry using the {\sc hstphot} package
(\citealt{2000PASP..112.1383D}).  The candidate counterpart present in
the summed images gives F336W$=23.27\pm0.15$, F555W$=23.92\pm0.08$,
F814W$=24.01\pm0.11$ ($U=23.7$, $V=23.9$, $I=24.0$). The error on the
reported photometry is mainly due to spread in aperture correction. As
the magnitudes in the $U$--, $V$--, and $I$--band depend on the
spectral type assumed when using {\sc synphot}, we do not provide
errors for their magnitudes.

Using {\sc hstphot} we placed artificial point sources of varying
magnitude in the region of 150$\times$150 pixels around the \chan\,
position to determine the limiting magnitude of each of the three
combined images separately.  We detect point sources at 5--$\sigma$
with F336W$<24.6$, F555W$<26.8$, F814W$< 25.7$ ($U<25.0$, $V<26.8$,
$I<25.7$).  The bright optical source is not resolved in the WFPC2
images although the sharpness parameter from {\sc hstphot} is at the
extreme end of the range for point sources pointing at a nearly
resolved source.

A second set of HST WFC2 observations was obtained on August 2, 2000
using the F300W filter (a wide $U$--band filter) and the F814W filter
for total exposure times of 1900~s and 260~s, respectively. Using {\sc
  hstphot} we find that the F300W magnitude of the source is
22.3$\pm$0.1, and the F814W magnitude is $23.5\pm0.1$. In both cases
the limiting magnitude of the exposures is $\approx 24$. See
Table~\ref{wfc3-phot} for the conversion of these F300W and F814W
magnitudes to Cousins $I$ and Johnson $U$--band magnitudes.

\subsection{{\it HST} WFC3 observations }

In addition to the WFPC2 HST observations, we have analysed archival HST
data obtained with the Wide Field Camera 3 (WFC3) in the UVIS F336W
and the F438W filters. The WFC3 UVIS detector has a pixel scale of
0.04\arcsec\, per pixel, which is a factor of 2.5 better than that of
WFPC2, and 0.13\arcsec\, per pixel in the infrared (IR) band.

The HST WFC3 data were obtained on Oct.~3, 2010 (MJD 55472) and the proposal
identification number is 11691.  We summed the reduced drizzled images
provided by the Space Telescope Science Institute giving total
exposure times of 2300~s and 950~s for the UVIS F336W and the F438W filters,
respectively. Furthermore, we analysed the observations obtained in
the near--infrared F110W-- (1565~s in total) and F160W--bands exposures
(2518~s in total).

\begin{figure*} 
\includegraphics[angle=0,width=16cm,clip]{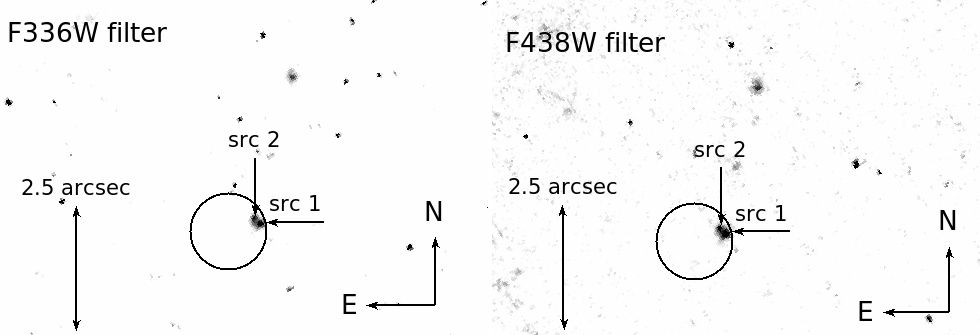}
\caption{North is up and East is to the left in both panels. {\it Left
    panel:} Zoom--in of the combined 2300~s WFC3 F336W image around
  the position of the bright ULX in \src. The circle of 0.76\arcsec\,
  radius represents the astrometric 90\% confidence region of the
  position of the ULX. {\it Right panel:} Zoom--in of the combined,
  950~s--long, HST WFC3 F438W image, showing that the source at the
  edge of the 0.76\arcsec\, radius 90\% confidence astrometric error
  circle is resolved into two sources.  }
\label{ima2} \end{figure*}

We used a procedure similar to the one we used for the WFPC2 images to
obtain an accurate astrometric solution for the WFC3 images. We
considered all the sources that are not saturated and appear stellar
and unblended. Using 11 stars on the F336W--band UVIS image and 7
stars on the near--infrared F110W image for the astrometry the fit for
the reference point position, the scale and the position angle yielded
rms residuals of 0.05\arcsec\, for the UVIS F336W--band image and
0.1\arcsec\, for the F110W near--infrared image. The astrometric
solution is against the SDSS catalogue for the F336W--band UVIS image and
against 2MASS for the near--infrared image. The bright point source
present inside the \chan\, error circle in the WFPC2 images is
resolved into two sources in the UVIS images with positions for star
1: RA=11:51:07.647 and Dec=+55:04:15.26 and for star 2:
RA=11:51:07.661 and Dec=+55:04:15.36 (star 1: [RA,Dec]=177.781862(14),
+55.070906(14); star 2: [RA,Dec]=177.781921(14), +55.070933(14) in
decimal degrees, where the value in the between brackets denotes the
1--$\sigma$ uncertainty; see Fig.~\ref{ima2}). The source is
unresolved in the near--infrared images. The combined 90\% astrometric
uncertainty of tying the HST/WFC3 images to the ICRS using the SDSS
images and of localizing the two stars on the WFC3 optical frames is
0.3\arcsec. The total \chan\, and HST/WFC3 combined 90\% astrometric
uncertainty is therefore 0.76\arcsec.

In deriving the photometry of the sources we used sky values
determined local to the source of interest as the sky background
varies substantially due to the presence of unresolved stars from the
galaxy. We performed point--spread function fitting for the photometry
using the {\sc dolphot} package. The WFC3 magnitudes in the {\sc
  vegamag} system for the resolved candidate optical counterpart are
given in Table~\ref{wfc3-phot}. The F336W magnitude of the two sources
combined is 23.29. A WFC3 F336W magnitude of 23.29 gives a magnitude
of 23.30 for colors similar to those of the O7~I star BZ~64 in the
WFPC2 F336W filter. This is fully consistent with the magnitude of
23.27$\pm$0.15 measured with WFPC2 in its F336W--band filter more than
16 year prior. The combined WFC3 F438W band magnitude of the two
sources is 23.43. Converting the WFC3 band magnitudes to Johnson $U$--
and $B$--band magnitudes again using the BZ~64 O7~I in {\sc synphot}
we get $U_1=24.3$ and $B_1=23.9$ and $U_2=24.7$ and $B_2=24.4$ for the
F336W and F438W magnitudes of the two sources separately, respectively
and $U=23.7$ and $B=23.4$ for the two sources combined.

\begin{table*}
  \caption{Results of the point--spread function photometry on the 
    candidate optical counterpart(s) in the astrometric error circle 
    using two HST
    WFPC2 ({\it top part}) and WFC3 ({\it bottom part}) observations. The 
    Johnson and
    Cousins magnitudes have been calculated using {\sc synphot} assuming a
    O7~I spectral type (the star BZ~64).}
\label{wfc3-phot}
\begin{center}
\begin{tabular}{ccccc}
\hline
Observing  & Source& Time on & HST Filter magnitude & Johnson
(UBV)/Cousins (I)\\
date & \# & source (s) & &  Magnitude\\
\hline
Jun.~28, 1994 & 1$+$2 & 1200 & F336W= 23.27$\pm$0.15& $U=23.7$\\
Jun.~28, 1994 & 1$+$2 & 2430 & F555W= 23.92$\pm$0.08& $V=23.9$\\
Jun.~28, 1994 & 1$+$2 & 1830 & F814W = 24.01$\pm$0.11& $I=24.0$\\
Aug.~8, 2000 & 1$+$2 & 1900 & F300W= 22.3$\pm$0.1& $U=22.9$\\
Aug.~8, 2000 & 1$+$2 & 260 & F814W = 23.5$\pm$0.1& $I=23.5$\\
\hline
Oct.~3, 2010 & 1 & 2300 & F336W = 23.85$\pm$0.03 & $U=24.3$\\
Oct.~3, 2010 & 2 & 2300 & F336W = 24.27$\pm$0.04 & $U=24.7$\\
Oct.~3, 2010 & 1$+$2 & 2300 & F336W = 23.29 & $U=23.7$\\
Oct.~3, 2010 & 1 & 950  & F438W = 23.97$\pm$0.05 & $B=23.9$\\
Oct.~3, 2010 & 2 & 950  & F438W = 24.45$\pm$0.07 & $B=24.4$\\
Oct.~3, 2010 & 1$+$2 & 950  & F438W = 23.43& $B=23.4$\\

\end{tabular}      
\end{center}
\end{table*}

In order to estimate the limiting magnitude of the WFC3 images we
checked the magnitudes of the faintest sources detected at 5--$\sigma$,
and we find approximate limiting magnitudes of F336W$<27.2$ and
F438W$<26.3$ ($U\approxlt27.6$ and $B\approxlt26.3$).

\section{VLA radio observations}

The field of \src\, was observed in the X--band at a central frequency
of 8.4 GHz with the Very Large Array (VLA) in A--configuration on two
occasions in 2003.  On June 19, 2003 the source was observed under
project code AN114.  The source 3C~48 was used as the primary
calibrator and the secondary calibrator was J~1146+539. In total the
observation provides 14 minutes on target.  On July 14/15, 2003 the
source was observed under project code AN115.  The source 3C~286 was
used as the primary calibrator and the secondary calibrator was
J~1148+529. In total the observation provides 25 minutes on target.

Data reduction was carried out according to standard procedures within
the Astronomical Image Processing System {\sc AIPS}. The field was not
bright enough for self-calibration.  When combined the two
observations gave a root--mean--square (rms) value of the noise of 29
$\mu$Jy/beam, and no detection at the position of the ULX down to a
3$\sigma$ upper limit to the radio flux of 87 $\mu$Jy.  This yields a
3$\sigma$ upper limit on the radio luminosity at the position of the
ULX of $5.6 \times 10^{36}$ \lum\, for a distance of 80 Mpc.

The only significantly detected source corresponds to the center of
the galaxy, at RA=11:51:06.8724(2) Dec=55:04:43.307(2) (J2000), at
~1.8 mJy. The errors on those positions are purely statistical.  The
VLA astrometry is thought to be good to about 0.1\arcsec\, in
reasonable weather conditions when in A--configuration. Therefore, we
have used 0.1\arcsec\, as the fiducial 1$-\sigma$ uncertainty on the
VLA coordinates of the AGN in \src.

\section{Discussion}

We report on a brief \chan\, observation of the field of the bright
ULX in \src\ (\citealt{2004MNRAS.353..221N}). Compared with the
discovery \xmm\, observation $\sim$9.5 years earlier, the source 0.5-10
keV X-ray luminosity has faded by a factor of $\sim$2.5. This virtually
proves that the high 0.5--10 keV X-ray luminosity of
2$\times10^{40}$\lum\, detected by \xmm\, is due to a single bright
source and not due to the chance super position of a group of
sources. If the latter were the case they, coincidentally, would have
decreased their X-ray luminosity by the time of the \chan\,
observation (cf.~\citealt{2007Natur.445..183M}).
 
Using the accurate position we obtained using \chan\, we investigate
archival HST WFPC2 images in the F300W, F336W, F555W and F814W
(2$\times$) bands (which we converted to the Johnson $U$--, $U$--,
$V$--, and Cousins $I$--bands, respectively) and archival WFC3 images
in the F336W and F438W bands (which we converted to the Johnson $U$--
and $B$--band magnitudes).  There is a single bright, blue optical
source in the astrometric error circle in the WFPC2 images. In the
WFC3 images this source is resolved into two sources that, when
combined, have the same F336W magnitude in the WFC3 and WFPC2
observations of 1994 and 2010 (but see below). With a
F336W$-$F438W=$-0.14$, $U-B=0.3$, and $U-V=-0.2$ color, this source is
blue (the color of source 1 and source 2 is approximately equally
blue).  Furthermore, if at the distance of \src, with a distance
modulus DM of 34.5 excluding extinction effects, its combined absolute
$V$-band magnitude, ${\rm M_V}$, is $-10.6$. The Galactic extinction
in the direction of \src\, is found to be $1.1\times 10^{20}$
cm$^{-2}$ (\citealt{1990ARA&A..28..215D}).  Using the relation of
\citet{1995A&A...293..889P} between ${\rm N_H}$ and ${\rm A_V}$ we
obtain ${\rm A_V}=0.06$. Using {\sc synphot} we calculate the
F336W-F438W color the O7~I star BZ~64 would have correcting for
extinction. The observed F336W-F438W color is too blue to be explained
by the O7~I star even if there was no extinction. The color of an O5~V
star (BZ~77) may be consistent with the observed F336W$-$F438W color
if the extinction is slightly lower than the E(B--V) of 0.18 derived
from \citet{1990ARA&A..28..215D}. So, we conclude that the source(s)
making up source 1 and 2 have a combined spectrum similar to that of
an early type star of approximately type O5.

A direct comparison of the two WFPC2 observations employing the F814W
filter obtained with just over 6 years in between provides evidence
for variability: in 1994 the F814W magnitude was 24.01$\pm$0.11,
whereas the observation in 2000 found F814W$=23.5\pm 0.1$. The \xmm\,
observation of \citet{2004MNRAS.353..221N} was obtained on April 27,
2002 when the source was at its highest X--ray luminosity detected so
far. The optical observation closest in time to the \xmm\, observation
also finds the source at a slightly enhanced luminosity, suggesting
that one of the two sources is indeed associated with the X--ray
source.  Furthermore, comparing the Johnson $U$--band magnitudes
calculated assuming an O7~I spectral type, the source is clearly
brighter in 2000 than in 1994 and 2010, with $U$--band magnitudes of
23.7, 22.9 and 23.7 for the observation in 1994, 2000, and 2010,
respectively, re-enforcing the evidence for variability.

The optical counterpart is classified as a stellar association or
large globular cluster by \citet{1996AJ....112.1839S} using the 1994
HST WFPC2 observations as we mentioned in Section~2.2 although they
did not consider the $U$-band observations.  In the WFC3 observations
the full--width at half maximum (FWHM) of source 1 and 2 is $\approx
2.4$ pixels. This yields a FWHM of $<$0.1\arcsec, which at the
distance of 80 Mpc of \src\, is about 40 pc for each source. This is
on the small side of the size distribution of OB associations
(cf.~\citealt{1998AJ....116..119B}) although the associations in
IC~1613 have sizes of 40 pc (\citealt{2010A&A...523A..23G}). The $U-V$
color is also consistent with that of young (10-50 Myr old) OB
associations (\citealt{2002A&A...392....1S};
\citealt{2003A&A...401.1063A}). In a few nearby ULXs, e.g.~
NGC~1313~X--2 and Holmberg~IX~X--1, the age of the surrounding stars
is found to be in the same range (\citealt{2008A&A...486..151G};
\citealt{2011ApJ...734...23G}). Similarly, assuming that one of the
optical sources is related to the ULX and the other is an OB cluster
the separation between them is similar to that for M~82~X--1 and
NGC~7479 found by \citet{2011MNRAS.418L.124V}.

Optical counterparts to ULXs with luminosities below $1\times
10^{40}$\lum\, are typically fainter than the absolute magnitude for
\src\,X--2, $M_V=-10.6$. The brightest have $M_V=-9$
(\citealt{2008MNRAS.387...73R}).  The ULXs with higher luminosities
are associated with brighter optical counterparts.  For instance, the
bright ULX in ESO 243-49 has $M_R=-10$ (\citealt{2010MNRAS.405..870S})
and the bright ULX in CXO~122518.6+144545 has $M_{g^\prime}=-10.1$
(\citealt{2010MNRAS.407..645J}). In the following we discuss the
possible nature of this bright ULX and the associated potential
counterpart.

\subsection{\src~X--2: a ULX with a stellar--mass black hole?}

The two candidate optical counterparts to \src~X--2 are equally blue in the WFC3
observation. Since the accretion disk color can be similarly blue
(e.g.~\citealt{1995xrb..book...58V}), perhaps the observed light of one of the
sources is dominated by the accretion disk. If so, the other source is dominated
by a cluster of stars of a combined color similar to that of the accretion
disk.  The observed X--ray and optical variability and in particular the large
variability in the $U$--band does suggest that one of the two optical sources
identified in the WFC3 images is indeed related to the counterpart to the ULX.
If one of the two sources is due to a single object, it could also be a Luminous
Blue Variable (LBV; \citealt{1994PASP..106.1025H}). If there is evidence that
this source is an LBV it is prudent to assume (Occam's razor) that it will be
the mass donor to the accreting black hole in the ULX. The reason is that both
LBVs and ULXs are so rare in a galaxy that the probability to find a ULX and an
LBV very close together while not physically related is the product of the two
(low) probabilities of finding one such source in a certain area in a galaxy.

On the other hand, one may be inclined to assume that evidence for an LBV for
one of the two optical sources implies that the other source must be the ULX,
reasoning that the number of LBVs in a binary orbit with a black hole must be
rarer still than single LBVs. However, most O stars are formed in a
binary with another (massive) star (\citealt{2012Sci...337..444S}), making the
scenario of finding an LBV with a black hole plausible.

Finally, a scenario where both blue optical sources detected in the
WFC3 images are young OB clusters is also valid as long as one of them
hosts the ULX. When the ULX is X--ray bright it has to have an optical
counterpart of magnitude similar to that of the OB association,
explaining the variability. Therefore, the scenario where the ULX in
\src~X--2 is explained as a stellar--mass black hole accreting from a
donor in a wide system such that the absolute magnitude of the donor
plus disk rivals that of an OB association is entirely consistent with
the data.

\subsection{\src~X--2: a type IIn supernova?}

As the combined $U$--band magnitude from the WFC3 source 1 and 2
measured in 2010 is consistent with being the same as the WFPC2
$U$--band magnitude measured in 1994 while there is evidence for
moderate brightening in 2000, a scenario explaining the ULX as a type
IIn supernova (cf.~\citealt{2009Natur.458..865G};
\citealt{2010MNRAS.407..645J}) is strongly constrained. The scenario
that still is (theoretically) possible but that we deem unlikely is as
follows: the progenitor star of the supernova was not detected,
therefore it should reside in a cluster that is still present. The
supernova went off some time between 1994 and 2000 and, remarkably,
went undetected in the optical. This includes a non-detection in
archival groundbased observations from the Jacobus Kaptein Telescope
in 1996. The supernova has faded in the optical before the HST
observation in 2000 nearly down to the pre-explosion level. The
X--rays come from the supernova and they are now fading.

\subsection{A new scenario for the nature of some ULXs. Tidal
  disruption of a star by a recoiled black hole?}

A galaxy like our Milky Way is predicted to have hundreds of recoiled
IMBHs/SMBHs with associated star clusters
(\citealt{2012MNRAS.421.2737O}).  For the majority of systems which
are expected to have had recoil velocities $<$1000 km s$^{-1}$, the
rate of tidal disruption events remains constant for about 1 Gyr after
the recoil event. Given the expected rate one could have about a 1000
events in 1 Gyr (\citealt{2008ApJ...683L..21K}).  Tidal disruption of
stars from such a bound (nuclear) cluster or from unbound stars of the
galaxy will produce periods of enhanced accretion providing X-ray
luminosities above $10^{40}$ \lum\, (\citealt{2008ApJ...683L..21K};
\citealt{2012MNRAS.422.1933S}). The short duration Eddington--limited
flares may well be missed by current scanning X-ray All Sky Monitors,
whereas the flares may be too soft to be detected by the {\it Swift}
BAT instrument (cf.~\citealt{2011MNRAS.410..359L} for the predicted
lightcurves and the disk temperatures).  Theoretically, one expects
the timescale of the decay of mass accretion rate to go as $t^{-5/3}$
(\citealt{1988Natur.333..523R}; \citealt{1990Sci...247..817R}). Recent
\chan\, observations of three potential tidal disruption events are
consistent with this theoretical trend and an overall lifetime of
approximately 5 years at L$_x >10^{40}$
\lum\,(\citealt{2004ApJ...604..572H}). In order to determine the
importance of tidal disruption events from recoiled, off-nuclear black
holes for the population of ULXs we need to estimate the number of
such black holes that recoiled in approximately the last Gyr. The rate
of tidal disruption events drops off steeply for longer times as the
loss-cone of the recoiled black hole is emptied
(\citealt{2008ApJ...683L..21K}).

\citet{2000ApJ...536..153P} found that over the last 1 Gyr about 1\%
of galaxies with an absolute $B$--band magnitude $-21 \leq {\rm M}_B
\leq -18$ went through a merger (cf.~their formula 32). This was
derived evaluating the number of interacting galaxies found in the
Second Southern Sky Redshift Survey. As the authors explain, this
fraction is a lower limit as only pairs of relatively bright galaxies
have been included. Including fainter galaxies would increase the
fraction as more interacting bright $+$ faint galaxy pairs would be
found. More recently, \citet{2007ApJ...666..212D} found a merger
fraction of 2\% using close pair fractions and asymmetries from the
Millenium Galaxy catalogue.

Assuming that the black hole merger time is significantly less than 1
Gyr such as in gas--rich mergers (e.g.~\citealt{2007Sci...316.1874M})
will provide a lower limit on the number of recently (within the last
Gyr) recoiled black holes. In the case that this merger time is much
longer (of order of several Gyrs) the number of recently recoiled
black holes is larger as the galaxy merger fraction was high at larger
redshift (e.g.~\citealt{2000ApJ...536..153P};
\citealt{2004ApJ...617L...9L}). 

In order to derive the number of off--nuclear recoiled black holes that recoiled
in approximately the last Gyr we need to know the fraction of merging black
holes that receive a significant kick. This fraction is debated at present. The
magnitude of the recoil kick depends on the mass ratio of the two black holes
before their merger, on the orientation of their spin axes with respect to the
binary orbital plane and on the amount of spin. The presence of massive gas
discs around the binary black hole such as potentially in gas--rich mergers can
align the spin axes of the two black holes prior to their merger, reducing the
amplitude of the recoil kick imparted during the merger
(\citealt{2007ApJ...661L.147B}; \citealt{2009MNRAS.tmp.1795D}). However,
fragmentation of the accretion disc around the binary SMBH could reduce the rate
of alignment as fragments could result in short randomly oriented accretion
events (\citealt{2007MNRAS.377L..25K}). Recent calculations seem to suggest that
the recoil kicks might be limited to velocities lower than $\approx 200$ km
s$^{-1}$ due to a gravitational wave "anti-kick" during the ringdown phase
(\citealt{2010CQGra..27a2001L}). Similarly, under certain conditions
relativistic precession will align or anti-align the black hole spin vectors
reducing the median kick velocities (\citealt{2010ApJ...715.1006K}).  Recoil
velocities of a few hundred km s$^{-1}$ imply that many recoiling SMBHs will be
retained by the host galaxy. Too small recoil velocities will mean that the
recoiled black hole will sink back to the nucleus of the galaxy and X-ray
emission due to decaying tidal disruption events will not be identified as ULXs.
For now we assume that 50\% of the recoiled black holes will have recoil
velocities such that they are bound to the galaxy and that they appear as
off-nuclear sources for 1 Gyr. We note however, that this number depends on the
distribution of recoil kick velocities which is uncertain.

Putting all these estimates together we calculate the probability of finding a
ULX caused by tidal disruption of a star by a recoiled SMBH per galaxy. There is
2\% (the merger fraction) $\times$ 50\% (right recoil velocity) $\times$
$1\times10^{-6}$/yr (the tidal disruption rate per year) $\times$ 5 (the
life--time in years above a luminosity of 1$\times 10^{40}$\lum) leading to
about 5 in $10^8$ galaxies that potentially has a bright off--nuclear X--ray
source due to the fading of emission caused by a tidal disruption event.

The 1 Gyr look--back time corresponds to a redshift of z$\approx
0.08$.  This provides a co-moving volume of 0.16 Gpc$^{-3}$. Given the
galaxy density $\phi_\star \approx 1\times 10^{-2}$ Mpc$^{-3}$
(cf.~\citealt{1976ApJ...203..297S}; \citealt{1992ApJ...390..338L};
\citealt{2005MNRAS.360...81D} for h=0.7) there are about 1.6 million
galaxies in this volume. Therefore, there will be between 0.1--1
X-ray source with L$>10^{40}$\lum\, due to tidally disrupted stars
from off-nuclear, recoiled, black holes within a redshift of 0.08.
Note that a luminosity of L$>10^{40}$\lum\, at a redshift of z=0.08
corresponds to a flux of only $\approx 7\times 10^{-16}$\flx\, making
it difficult to detect these sources in the existing shallow
surveys. If the estimates above are reasonable there might
be at most one of these among the (bright) ULXs such as the ULX
\src~X--2.

\subsubsection{Application to \src~X--2}

The decay in X--ray luminosity of a factor of 2.5 in \src~X--2 that we find
between the \xmm\, and \chan\, observations $\sim3400$ days apart is (too) small
compared to the predicted drop-off from the soft disk X--ray spectral component
predicted by \citet{2011MNRAS.410..359L}. Even if in some proposed tidal
disruption events the X--ray spectrum is found to be significantly harder than
expected, the luminosity is about a factor 200 lower 3165~days after the soft
peak (\citealt{2004ApJ...603L..17K}). Given the presence of nuclear activity in
\src\, a new central SMBH must have replaced the recoiled black hole, for
instance following a recent merger, or the recoiled black hole was formed by a
three black hole interaction event. All these constraints of course make the
scenario of a tidal disruption event in a recoiled SMBH for \src\,X--2 unlikely.

\subsection{Alternative explanations for \src~X--2?}

Alternative explanations for the bright ULX in \src\, such as a background AGN
or a foreground star or (quiescent) low--mass X--ray binary (LMXB) are possible
in principle. For \src~X-2 to be an active LMXB the distance should be
$\approx$1 Mpc for a luminosity of 1$\times 10^{36}$\lum\, typical for an LMXB.
This would give an absolute $V$--band magnitude of around -1 which is in line
with that of LMXBs (\citealt{1995xrb..book...58V}), however, finding an LMXB in
interstellar space is unlikely. \src~X--2 can also be a quiescent LMXB at a
distance of a few kpc. The absolute magnitude of the optical counterpart would
then be pointing to an M--type companion, which would imply that the blue
candidate optical counterparts have to be unrelated to the quiescent LMXB.

\citet{1996AJ....112.1839S} estimate the number of background galaxies
in the WFPC2 field of view as 1.2$\times 10^{-3}$ per
arcsec$^{-2}$. The presence of the background galaxy cluster Abell
1400 is taken into account in this estimate. The optical counterpart
does coincide spatially with one of the tidal tails that clearly
belongs to \src. As the (young) stellar associations in \src\, trace
these tails very well the probability of finding a background AGN
among these stellar associations is low.  

In the case that the optical source is unrelated to the X-ray source
the observed variability in the F814W (and $U$--band) WFPC2
observations is hard to explain.  Nevertheless, assuming the
variability is spurious or explained in another way, we can provide a
limit on the absolute magnitude of the counterpart of M$_U>-7.3$,
M$_B>-8.2$, M$_V>-7.7$ and M$_I>-8.8$. These limits on M$_U$, M$_B$,
M$_V$ and M$_I$ have been derived using the upper limits on the
apparent magnitude of $<27.2$, $<26.3$, $<26.8$ and $<25.7$,
respectively and given the distance of 80 Mpc towards \src.

Therefore, the source could be a ULX with a stellar
counterpart outside an association.  The upper limit to an optical
source in this scenario translates to a lower limit on the ratio of
the X--ray flux to the optical flux of 12.  Whereas most
\chan--selected AGN have a ratio of X--ray to optical $R$--band flux
of lower than 10 (\citealt{2009ApJS..180..102L}), our limit does not
rule out a background AGN scenario in the case that the optical source
is unrelated to the X--ray source. The most secure way to differentiate
between the discussed scenarios is to obtain an optical spectrum of
the potential counterparts.

\section*{Acknowledgments} \noindent PGJ and MAPT acknowledge support from the Netherlands
Organisation for Scientific Research. PGJ acknowledges discussions with Elena Rossi, Suvi
Gezari and Luca Zampieri on tidal disruption flares. PGJ further acknowledges Rasmus Voss for
his comments on an earlier version of the manuscript. MH, PGJ and MAPT acknowledge input from
Dolphin on the HST photometry. GM thanks the Spanish Ministry of Science and Innovation (now
MINECO)  for support through grant AYA2010-21490-C02-02. This research used the HST Archive
facilities of the STScI, the ST-ECF and the CADC with the support of the following granting
agencies: NASA/NSF, ESA, NRC, CSA. This research has made use of the SIMBAD database, operated
at CDS, Strasbourg, France, of NASA's Astrophysics Data System Bibliographic Services, of
SAOImage DS9, developed by Smithsonian Astrophysical Observatory, of software provided by the
Chandra X-ray Center (CXC) in the application packages CIAO, and of the XRT Data Analysis
Software (XRTDAS) developed under the responsibility of the ASI Science Data Center (ASDC),
Italy.


\begin{thebibliography}{67}
\expandafter\ifx\csname natexlab\endcsname\relax\def\natexlab#1{#1}\fi

\bibitem[{{Adelman-McCarthy}(2009)}]{2009yCat.2294....0A}
{Adelman-McCarthy}, J.~K.~e., 2009, VizieR Online Data Catalog, 2294, 0

\bibitem[{{Anders} \& {Fritze-v.~Alvensleben}(2003)}]{2003A&A...401.1063A}
{Anders}, P., {Fritze-v.~Alvensleben}, U., 2003, \aap, 401, 1063

\bibitem[{{Begelman}(2002)}]{2002ApJ...568L..97B}
{Begelman}, M.~C., 2002, \apjl, 568, L97

\bibitem[{{Bogdanovi{\'c}} et~al.(2007){Bogdanovi{\'c}}, {Reynolds}, \&
  {Miller}}]{2007ApJ...661L.147B}
{Bogdanovi{\'c}}, T., {Reynolds}, C.~S., {Miller}, M.~C., 2007, \apjl, 661,
  L147

\bibitem[{{Bresolin} et~al.(1998)}]{1998AJ....116..119B}
{Bresolin}, F., et~al., 1998, \aj, 116, 119

\bibitem[{{Colbert} \& {Mushotzky}(1999)}]{1999ApJ...519...89C}
{Colbert}, E.~J.~M., {Mushotzky}, R.~F., 1999, \apj, 519, 89

\bibitem[{{De Propris} et~al.(2007){De Propris}, {Conselice}, {Liske},
  {Driver}, {Patton}, {Graham}, \& {Allen}}]{2007ApJ...666..212D}
{De Propris}, R., {Conselice}, C.~J., {Liske}, J., {Driver}, S.~P., {Patton},
  D.~R., {Graham}, A.~W., {Allen}, P.~D., 2007, \apj, 666, 212

\bibitem[{{Dickey} \& {Lockman}(1990)}]{1990ARA&A..28..215D}
{Dickey}, J.~M., {Lockman}, F.~J., 1990, \araa, 28, 215

\bibitem[{{Dolphin}(2000)}]{2000PASP..112.1383D}
{Dolphin}, A.~E., 2000, \pasp, 112, 1383

\bibitem[{{Dotti} et~al.(2009){Dotti}, {Volonteri}, {Perego}, {Colpi},
  {Ruszkowski}, \& {Haardt}}]{2009MNRAS.tmp.1795D}
{Dotti}, M., {Volonteri}, M., {Perego}, A., {Colpi}, M., {Ruszkowski}, M.,
  {Haardt}, F., 2009, \mnras, 1795

\bibitem[{{Driver} et~al.(2005){Driver}, {Liske}, {Cross}, {De Propris}, \&
  {Allen}}]{2005MNRAS.360...81D}
{Driver}, S.~P., {Liske}, J., {Cross}, N.~J.~G., {De Propris}, R., {Allen},
  P.~D., 2005, \mnras, 360, 81

\bibitem[{{Farrell} et~al.(2009){Farrell}, {Webb}, {Barret}, {Godet}, \&
  {Rodrigues}}]{2009Natur.460...73F}
{Farrell}, S.~A., {Webb}, N.~A., {Barret}, D., {Godet}, O., {Rodrigues}, J.~M.,
  2009, \nat, 460, 73

\bibitem[{{Feng} \& {Kaaret}(2009)}]{2009ApJ...696.1712F}
{Feng}, H., {Kaaret}, P., 2009, \apj, 696, 1712

\bibitem[{{Feng} \& {Soria}(2011)}]{2011NewAR..55..166F}
{Feng}, H., {Soria}, R., 2011, New Astronomy Review, 55, 166

\bibitem[{{Gal-Yam} \& {Leonard}(2009)}]{2009Natur.458..865G}
{Gal-Yam}, A., {Leonard}, D.~C., 2009, \nat, 458, 865

\bibitem[{{Garcia} et~al.(2010){Garcia}, {Herrero}, {Castro}, {Corral}, \&
  {Rosenberg}}]{2010A&A...523A..23G}
{Garcia}, M., {Herrero}, A., {Castro}, N., {Corral}, L., {Rosenberg}, A., 2010,
  \aap, 523, A23

\bibitem[{{Garmire}(1997)}]{1997AAS...190.3404G}
{Garmire}, G.~P., 1997, in American Astronomical Society Meeting Abstracts
  \#190, vol.~29 of \emph{Bulletin of the American Astronomical Society}, p.
  823

\bibitem[{{Gladstone} et~al.(2009){Gladstone}, {Roberts}, \&
  {Done}}]{2009MNRAS.397.1836G}
{Gladstone}, J.~C., {Roberts}, T.~P., {Done}, C., 2009, \mnras, 397, 1836

\bibitem[{{Gris{\'e}} et~al.(2008){Gris{\'e}}, {Pakull}, {Soria}, {Motch},
  {Smith}, {Ryder}, \& {B{\"o}ttcher}}]{2008A&A...486..151G}
{Gris{\'e}}, F., {Pakull}, M.~W., {Soria}, R., {Motch}, C., {Smith}, I.~A.,
  {Ryder}, S.~D., {B{\"o}ttcher}, M., 2008, \aap, 486, 151

\bibitem[{{Gris{\'e}} et~al.(2011){Gris{\'e}}, {Kaaret}, {Pakull}, \&
  {Motch}}]{2011ApJ...734...23G}
{Gris{\'e}}, F., {Kaaret}, P., {Pakull}, M.~W., {Motch}, C., 2011, \apj, 734,
  23

\bibitem[{{Halpern} et~al.(2004){Halpern}, {Gezari}, \&
  {Komossa}}]{2004ApJ...604..572H}
{Halpern}, J.~P., {Gezari}, S., {Komossa}, S., 2004, \apj, 604, 572

\bibitem[{{Hong} et~al.(2005){Hong}, {van den Berg}, {Schlegel}, {Grindlay},
  {Koenig}, {Laycock}, \& {Zhao}}]{2005ApJ...635..907H}
{Hong}, J., {van den Berg}, M., {Schlegel}, E.~M., {Grindlay}, J.~E., {Koenig},
  X., {Laycock}, S., {Zhao}, P., 2005, \apj, 635, 907

\bibitem[{{Humphreys} \& {Davidson}(1994)}]{1994PASP..106.1025H}
{Humphreys}, R.~M., {Davidson}, K., 1994, \pasp, 106, 1025

\bibitem[{{Jonker} et~al.(2010){Jonker}, {Torres}, {Fabian}, {Heida},
  {Miniutti}, \& {Pooley}}]{2010MNRAS.407..645J}
{Jonker}, P.~G., {Torres}, M.~A.~P., {Fabian}, A.~C., {Heida}, M., {Miniutti},
  G., {Pooley}, D., 2010, \mnras, 407, 645

\bibitem[{{Kesden} et~al.(2010){Kesden}, {Sperhake}, \&
  {Berti}}]{2010ApJ...715.1006K}
{Kesden}, M., {Sperhake}, U., {Berti}, E., 2010, \apj, 715, 1006

\bibitem[{{King} \& {Pringle}(2007)}]{2007MNRAS.377L..25K}
{King}, A.~R., {Pringle}, J.~E., 2007, \mnras, 377, L25

\bibitem[{{King} et~al.(2001){King}, {Davies}, {Ward}, {Fabbiano}, \&
  {Elvis}}]{2001ApJ...552L.109K}
{King}, A.~R., {Davies}, M.~B., {Ward}, M.~J., {Fabbiano}, G., {Elvis}, M.,
  2001, \apjl, 552, L109

\bibitem[{{Komossa} \& {Merritt}(2008)}]{2008ApJ...683L..21K}
{Komossa}, S., {Merritt}, D., 2008, \apjl, 683, L21

\bibitem[{{Komossa} et~al.(2004){Komossa}, {Halpern}, {Schartel}, {Hasinger},
  {Santos-Lleo}, \& {Predehl}}]{2004ApJ...603L..17K}
{Komossa}, S., {Halpern}, J., {Schartel}, N., {Hasinger}, G., {Santos-Lleo},
  M., {Predehl}, P., 2004, \apjl, 603, L17

\bibitem[{{Laird} et~al.(2009)}]{2009ApJS..180..102L}
{Laird}, E.~S., et~al., 2009, \apjs, 180, 102

\bibitem[{{Le Tiec} et~al.(2010){Le Tiec}, {Blanchet}, \&
  {Will}}]{2010CQGra..27a2001L}
{Le Tiec}, A., {Blanchet}, L., {Will}, C.~M., 2010, Classical and Quantum
  Gravity, 27, 012001

\bibitem[{{Li} et~al.(2004){Li}, {Kastner}, {Prigozhin}, {Schulz}, {Feigelson},
  \& {Getman}}]{2004ApJ...610.1204L}
{Li}, J., {Kastner}, J.~H., {Prigozhin}, G.~Y., {Schulz}, N.~S., {Feigelson},
  E.~D., {Getman}, K.~V., 2004, \apj, 610, 1204

\bibitem[{{Li} et~al.(2008){Li}, {Li}, {Wang}, {Irwin}, \&
  {Rossa}}]{2008MNRAS.390...59L}
{Li}, J., {Li}, Z., {Wang}, Q.~D., {Irwin}, J.~A., {Rossa}, J., 2008, \mnras,
  390, 59

\bibitem[{{Lin} et~al.(2004)}]{2004ApJ...617L...9L}
{Lin}, L., et~al., 2004, \apjl, 617, L9

\bibitem[{{Lodato} \& {Rossi}(2011)}]{2011MNRAS.410..359L}
{Lodato}, G., {Rossi}, E.~M., 2011, \mnras, 410, 359

\bibitem[{{Loveday} et~al.(1992){Loveday}, {Peterson}, {Efstathiou}, \&
  {Maddox}}]{1992ApJ...390..338L}
{Loveday}, J., {Peterson}, B.~A., {Efstathiou}, G., {Maddox}, S.~J., 1992,
  \apj, 390, 338

\bibitem[{{Maccarone} et~al.(2007){Maccarone}, {Kundu}, {Zepf}, \&
  {Rhode}}]{2007Natur.445..183M}
{Maccarone}, T.~J., {Kundu}, A., {Zepf}, S.~E., {Rhode}, K.~L., 2007, \nat,
  445, 183

\bibitem[{{Mayer} et~al.(2007){Mayer}, {Kazantzidis}, {Madau}, {Colpi},
  {Quinn}, \& {Wadsley}}]{2007Sci...316.1874M}
{Mayer}, L., {Kazantzidis}, S., {Madau}, P., {Colpi}, M., {Quinn}, T.,
  {Wadsley}, J., 2007, Science, 316, 1874

\bibitem[{{Mineo} et~al.(2012){Mineo}, {Gilfanov}, \&
  {Sunyaev}}]{2012MNRAS.419.2095M}
{Mineo}, S., {Gilfanov}, M., {Sunyaev}, R., 2012, \mnras, 419, 2095

\bibitem[{{Miyawaki} et~al.(2009){Miyawaki}, {Makishima}, {Yamada}, {Gandhi},
  {Mizuno}, {Kubota}, {Tsuru}, \& {Matsumoto}}]{2009PASJ...61S.263M}
{Miyawaki}, R., {Makishima}, K., {Yamada}, S., {Gandhi}, P., {Mizuno}, T.,
  {Kubota}, A., {Tsuru}, T.~G., {Matsumoto}, H., 2009, \pasj, 61, 263

\bibitem[{{Nolan} et~al.(2004){Nolan}, {Ponman}, {Read}, \&
  {Schweizer}}]{2004MNRAS.353..221N}
{Nolan}, L.~A., {Ponman}, T.~J., {Read}, A.~M., {Schweizer}, F., 2004, \mnras,
  353, 221

\bibitem[{{Okajima} et~al.(2006){Okajima}, {Ebisawa}, \&
  {Kawaguchi}}]{2006ApJ...652L.105O}
{Okajima}, T., {Ebisawa}, K., {Kawaguchi}, T., 2006, \apjl, 652, L105

\bibitem[{{O'Leary} \& {Loeb}(2012)}]{2012MNRAS.421.2737O}
{O'Leary}, R.~M., {Loeb}, A., 2012, \mnras, 421, 2737

\bibitem[{{Patton} et~al.(2000){Patton}, {Carlberg}, {Marzke}, {Pritchet}, {da
  Costa}, \& {Pellegrini}}]{2000ApJ...536..153P}
{Patton}, D.~R., {Carlberg}, R.~G., {Marzke}, R.~O., {Pritchet}, C.~J., {da
  Costa}, L.~N., {Pellegrini}, P.~S., 2000, \apj, 536, 153

\bibitem[{{Pier} et~al.(2003){Pier}, {Munn}, {Hindsley}, {Hennessy}, {Kent},
  {Lupton}, \& {Ivezi{\'c}}}]{2003AJ....125.1559P}
{Pier}, J.~R., {Munn}, J.~A., {Hindsley}, R.~B., {Hennessy}, G.~S., {Kent},
  S.~M., {Lupton}, R.~H., {Ivezi{\'c}}, {\v Z}., 2003, \aj, 125, 1559

\bibitem[{{Predehl} \& {Schmitt}(1995)}]{1995A&A...293..889P}
{Predehl}, P., {Schmitt}, J.~H.~M.~M., 1995, \aap, 293, 889

\bibitem[{{Read} et~al.(2011){Read}, {Rosen}, {Saxton}, \&
  {Ramirez}}]{2011A&A...534A..34R}
{Read}, A.~M., {Rosen}, S.~R., {Saxton}, R.~D., {Ramirez}, J., 2011, \aap, 534,
  A34

\bibitem[{{Rees}(1988)}]{1988Natur.333..523R}
{Rees}, M.~J., 1988, \nat, 333, 523

\bibitem[{{Rees}(1990)}]{1990Sci...247..817R}
{Rees}, M.~J., 1990, Science, 247, 817

\bibitem[{{Roberts} et~al.(2008){Roberts}, {Levan}, \&
  {Goad}}]{2008MNRAS.387...73R}
{Roberts}, T.~P., {Levan}, A.~J., {Goad}, M.~R., 2008, \mnras, 387, 73

\bibitem[{{Sana} et~al.(2012)}]{2012Sci...337..444S}
{Sana}, H., et~al., 2012, Science, 337, 444

\bibitem[{{Schechter}(1976)}]{1976ApJ...203..297S}
{Schechter}, P., 1976, \apj, 203, 297

\bibitem[{{Schulz} et~al.(2002){Schulz}, {Fritze-v.~Alvensleben}, {M{\"o}ller},
  \& {Fricke}}]{2002A&A...392....1S}
{Schulz}, J., {Fritze-v.~Alvensleben}, U., {M{\"o}ller}, C.~S., {Fricke},
  K.~J., 2002, \aap, 392, 1

\bibitem[{{Schweizer}(1996)}]{1996AJ....111..109S}
{Schweizer}, F., 1996, \aj, 111, 109

\bibitem[{{Schweizer} et~al.(1996){Schweizer}, {Miller}, {Whitmore}, \&
  {Fall}}]{1996AJ....112.1839S}
{Schweizer}, F., {Miller}, B.~W., {Whitmore}, B.~C., {Fall}, S.~M., 1996, \aj,
  112, 1839

\bibitem[{{Soria} et~al.(2010){Soria}, {Hau}, {Graham}, {Kong}, {Kuin}, {Li},
  {Liu}, \& {Wu}}]{2010MNRAS.405..870S}
{Soria}, R., {Hau}, G.~K.~T., {Graham}, A.~W., {Kong}, A.~K.~H., {Kuin},
  N.~P.~M., {Li}, I.-H., {Liu}, J.-F., {Wu}, K., 2010, \mnras, 405, 870

\bibitem[{{Stone} \& {Loeb}(2012)}]{2012MNRAS.422.1933S}
{Stone}, N., {Loeb}, A., 2012, \mnras, 422, 1933

\bibitem[{{Strohmayer} \& {Mushotzky}(2003)}]{2003ApJ...586L..61S}
{Strohmayer}, T.~E., {Mushotzky}, R.~F., 2003, \apjl, 586, L61

\bibitem[{{Sutton} et~al.(2012){Sutton}, {Roberts}, {Walton}, {Gladstone}, \&
  {Scott}}]{2012MNRAS.423.1154S}
{Sutton}, A.~D., {Roberts}, T.~P., {Walton}, D.~J., {Gladstone}, J.~C.,
  {Scott}, A.~E., 2012, \mnras, 423, 1154

\bibitem[{{Swartz} et~al.(2011){Swartz}, {Soria}, {Tennant}, \&
  {Yukita}}]{2011ApJ...741...49S}
{Swartz}, D.~A., {Soria}, R., {Tennant}, A.~F., {Yukita}, M., 2011, \apj, 741,
  49

\bibitem[{{van Paradijs} \& {McClintock}(1995)}]{1995xrb..book...58V}
{van Paradijs}, J., {McClintock}, J.~E., 1995, p.~58 in X-ray Binaries,
  eds.~W.H.G.~Lewin, J.~van Paradijs, and E.P.J.~van den Heuvel, Cambridge:
  Cambridge Univ.~Press

\bibitem[{{Volonteri}(2010)}]{2010Natur.466.1049V}
{Volonteri}, M., 2010, \nat, 466, 1049

\bibitem[{{Voss} et~al.(2011){Voss}, {Nielsen}, {Nelemans}, {Fraser}, \&
  {Smartt}}]{2011MNRAS.418L.124V}
{Voss}, R., {Nielsen}, M.~T.~B., {Nelemans}, G., {Fraser}, M., {Smartt}, S.~J.,
  2011, \mnras, 418, L124

\bibitem[{{Walton} et~al.(2011){Walton}, {Roberts}, {Mateos}, \&
  {Heard}}]{2011MNRAS.416.1844W}
{Walton}, D.~J., {Roberts}, T.~P., {Mateos}, S., {Heard}, V., 2011, \mnras,
  416, 1844

\bibitem[{{Weisskopf} et~al.(2002){Weisskopf}, {Brinkman}, {Canizares},
  {Garmire}, {Murray}, \& {Van Speybroeck}}]{2002PASP..114....1W}
{Weisskopf}, M.~C., {Brinkman}, B., {Canizares}, C., {Garmire}, G., {Murray},
  S., {Van Speybroeck}, L.~P., 2002, \pasp, 114, 1

\bibitem[{{Wiersema} et~al.(2010){Wiersema}, {Farrell}, {Webb}, {Servillat},
  {Maccarone}, {Barret}, \& {Godet}}]{2010ApJ...721L.102W}
{Wiersema}, K., {Farrell}, S.~A., {Webb}, N.~A., {Servillat}, M., {Maccarone},
  T.~J., {Barret}, D., {Godet}, O., 2010, \apjl, 721, L102

\bibitem[{{Wolter} et~al.(2006){Wolter}, {Trinchieri}, \&
  {Colpi}}]{2006MNRAS.373.1627W}
{Wolter}, A., {Trinchieri}, G., {Colpi}, M., 2006, \mnras, 373, 1627

\end{thebibliography}
\end{document}